\documentclass[runningheads]{llncs}
\usepackage[T1]{fontenc}
\usepackage{graphicx}
\usepackage{booktabs} 
\usepackage{multirow}
\usepackage{amsmath}
\usepackage{multibib}
\usepackage{algorithm}
\usepackage{algpseudocode}
\usepackage{hyperref}
\usepackage[table,xcdraw]{xcolor}
\usepackage{multirow}
\usepackage{tabularx} 
\usepackage{xcolor}
\usepackage[table]{xcolor}
\usepackage{float}
\usepackage{gensymb}
\usepackage{notoccite}
\usepackage{authblk}
\usepackage{xcolor}
\usepackage{siunitx}
\usepackage{parskip}
\title{Two-Stage nnU-Net for Automatic Multi-class Bi-Atrial Segmentation from LGE-MRIs}

\author{Y. On \inst{1} \and
C. Galazis \inst{2} \and
C. Chiu \inst{3} \and
M. Varela \inst{1,4}
}
\date{September 2024}

\authorrunning{On et al.}
\titlerunning{Multi-class Bi-Atrial Segmentation Network}

\institute{National Heart \& Lung Institute, Imperial College London, UK \and Department of Computing, Imperial College London, UK \and Department of Electrical \& Electronic Engineering, Imperial College London, UK \and Cardiovascular \& Genomics Research Institute, City St George's University of London, UK \newline
\email{yu.on16@imperial.ac.uk}}

\begin{document}

\maketitle

\begin{abstract}
Late gadolinium enhancement magnetic resonance imaging (LGE-MRI) is used to visualise atrial fibrosis and scars, providing important information for personalised atrial fibrillation (AF) treatments. Since manual analysis and delineations of these images can be both labour-intensive and subject to variability, we develop an automatic pipeline to perform segmentation of the left atrial (LA) cavity, the right atrial (RA) cavity, and the wall of both atria on LGE-MRI. Our method is based on a two-stage nnU-Net architecture, combining 2D and 3D convolutional networks, and incorporates adaptive histogram equalisation to improve tissue contrast in the input images and morphological operations on the output segmentation maps. We achieve Dice similarity coefficients of $0.92\pm0.03$, $0.93\pm0.03$, $0.71\pm0.05$ and $95\%$ Hausdorff distances of \SI{3.89\pm6.67}{\mm}, \SI{4.42\pm1.66}{\mm} and \SI{3.94\pm1.83}{\mm} for LA, RA, and wall, respectively. The accurate delineation of the LA, RA and the myocardial wall is the first step in analysing atrial structure in cardiovascular patients, especially those with AF. This can allow clinicians to provide adequate and personalised treatment plans in a timely manner.


\end{abstract}

\section{Introduction}
Atrial fibrillation (AF) is characterised by irregular and rapid electrical activity originating from the atrial myocardium. It is the most common form of cardiac arrhythmia, affecting around 37 million people worldwide~\cite{Lippi2021GlobalChallenge}. One common treatment option is catheter ablation, in which specific areas of the heart are ablated (scarred) to block the propagation of abnormal electrical signals. There is some evidence that the degree and distribution of atrial fibrosis is an important determinant of AF treatment success~\cite{Li2022MedicalReview}. Moreover, it has been suggested that ablating atrial fibrotic regions may lead to better outcomes in some AF patients~\cite{marrouche2022effect}. 

Late gadolinium enhancement (LGE) is a technique commonly used in cardiac magnetic resonance imaging (MRI) to study myocardial scars and fibrosis. Due to the different washout kinetics of the administered gadolinium-based contrast agents, regions of myocardial injury show enhanced intensities compared to healthy tissues >10 minutes after administration~\cite{Li2022MedicalReview}. LGE-MRI has been proposed as a useful tool for AF treatment planning. However, clinical studies are mostly based on time-consuming manual analysis, and reliable automatic methods for LGE-MRI analysis are highly desired.

To this end, we introduce an automatic two-stage segmentation method for LGE-MRI of the atria, as part of the 2024 Multi-class Bi-Atrial Segmentation (MBAS) Challenge under MICCAI.

\section{Methods}
\begin{figure}[htb!]
\centering
\includegraphics[width=1.0\textwidth]{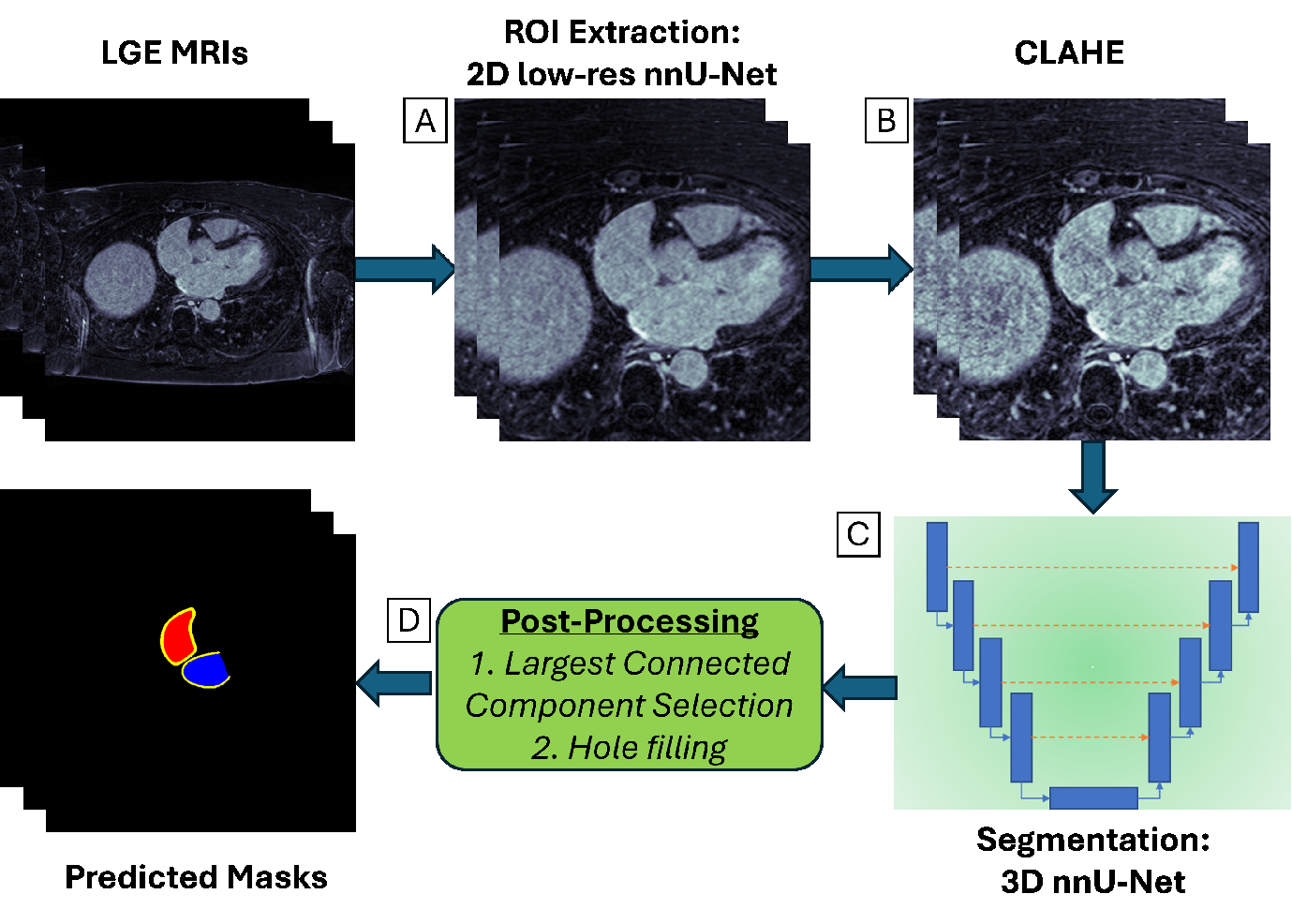}
\caption{The proposed segmentation pipeline. 3D LGE-MRIs are inputs to a \textbf{(A)} 2D nnU-Net to extract a region of interest (ROI) encapsulating the heart. \textbf{(B)} CLAHE is applied to the ROIs to improve the contrast of the images, which are the inputs to a \textbf{(C)} 3D nnU-Net that segments the LA, RA, and myocardial wall. The predicted masks are \textbf{(D)} post-processed with largest connected component selection and hole filling. \textit{CLAHE: Contrast limited adaptive histogram equalisation; LA: Left atrium; RA: right atrium}}\label{pipeline}
\end{figure}

The proposed segmentation pipeline consists of four main steps: region of interest (ROI) extraction, image histogram equalisation, segmentation, and morphological post-processing, as illustrated in Fig.~\ref{pipeline}. ROI extraction is performed in the first stage using a lightweight 2D nnU-Net. In the second stage, a 3D nnU-Net is used for segmentation. Aligning with the competition's rules, all neural networks are trained on the provided dataset and no pre-trained weights are used.

\subsection{Data}
The multi-centre dataset provided consists of 70 3D bi-atrial LGE-MRI scans for training, 30 for validation, and 100 for the final test phase of the competition. The training set comes with ground truth (GT) segmentation masks for three classes: left atrial (LA) cavity, right atrial (RA) cavity, and left and right atrial wall. The images have a voxel spacing of $0.625\times0.625\times2.5$~\unit{\mm^3}  and an image matrix size of either $640\times640\times44$ or $576\times576\times44$. While the test set images maintain the same voxel spacing, they may have different matrix sizes. This dataset expands upon a previous challenge~\cite{Xiong2021AImaging}, which focused exclusively on the LA, with the ultimate goal of improving ablation outcomes for patients with AF.

\subsection{Pre-Processing}
\subsubsection{Model Architecture: Image Cropping}
To reduce computational complexity, standardise image size, and minimise runtime in segmenting the LA and RA, we crop the images to include only the heart and surrounding structures. The steps we take are outlined below.

First, we merge the three original segmentation masks into a single binary image. Then, the images are downsampled using linear interpolation, with an in-plane reduction factor of 4 and a through-plane reduction factor of 2, to reduce inference time. The segmentation maps are similarly downsampled using a nearest-neighbour scheme.

We use a modified three-level 2D convolutional nnU-Net~\cite{Isensee2021NnU-Net:Segmentation} (version 2.4.1) implemented in PyTorch, for slices acquired perpendicular to the long axis. The convolutional layers use a feature map size of 16, 32, and 64 at each level, a $3\times3$ kernel with strides of $1\times 1$, $2\times 2$, and $2\times 2$, respectively, resulting in a 120k-parameter model. All other hyperparameters are retained as set by the default nnU-Net configuration. The network is optimised for 250 epochs. For inference, we select the weights that result in the best validation loss, which is the combined Dice and cross-entropy loss.

For ROI extraction, we calculate the 3D centroid on the predicted binary segmentation map and crop the image using a $320\times320\times44$ voxel box, ensuring surrounding context is preserved. Voxels extending beyond the image boundaries are padded with a value of 0.

\subsubsection{Contrast Limited Adaptive Histogram Equalisation}
Even for human experts, it can be difficult to delineate the myocardium wall from the LA or RA in poor-contrast regions. To tackle this problem, Contrast Limited Adaptive Histogram Equalisation (CLAHE)~\cite{Reza2004RealizationEnhancement,Stimper2019MultidimensionalEqualization} is applied as a pre-processing step to improve tissue contrast in the images. Unlike global histogram equalisation~\cite{Abdullah-Al-Wadud2007AEnhancement}, which calculates the histogram of pixel intensities and performs equalisation globally on the cumulative density function, CLAHE applies the equalisation on a local scale. Briefly, the input image is divided into a $H \times W \times D$ grid, in which the size of $H/W/D$ is $1/8$ of the corresponding dimension of the image, and the histogram is calculated separately in each grid.

\subsection{Segmentation}
\subsubsection{Model Architecture: Segmentation}
To segment the images into the three atrial categories, we use the 3D convolutional nnU-Net model~\cite{Isensee2021NnU-Net:Segmentation} with a total of 3M trainable parameters. To improve generalisability, we apply data augmentations during training, which consist of spatial transformations (e.g. rotation, transpose, mirroring), blurring, sharpening, additive white noise, brightening, and gamma correction. Details on specific parameters can be found in the nnU-Net's ``nnUNetTrainerDA5'' configuration. The model is optimised for 1,000 epochs. The predicted segmentation maps are then resized back to their original dimensions using nearest-neighbour resampling.


%

\subsection{Post-Processing}
To improve the quality of the segmentation maps, two morphological operations are employed as a post-processing step using SimpleITK~\cite{Lowekamp2013TheSimpleITK}: largest connected component selection and greyscale hole filling. Both operations are configured for a 3D face connectivity of 6. 

\subsection{Experimental Setup}
We evaluate the models using the Dice similarity coefficient (DSC) and 95\% Hausdorff distance (HD95, in \unit{\mm}). The HD95 represents the HD calculated based on the closest 95\% of points between two sets. During the development phase, we perform a 5-fold validation on the training set. For the submission to the validation and test phases of the competition, we optimise the weights by randomly splitting the 70 training cases into 65 for training and 5 validation.

The neural networks are trained on a single Nvidia RTX 6000 GPU, and the remaining pre-processing and post-processing steps are executed on a 3XS Intel Core i7 (10700K, 3.8GHz, 8 Core) CPU. The approximate inference time for a single case is 19.5~s. This includes 7.7~s, 0.5~s, 11~s, and 0.3~s, for ROI extraction, CLAHE, segmentation, and post-processing steps, respectively.

\section{Results}
\begin{figure}[ht!]
\centering
\includegraphics[width=1.0\textwidth]{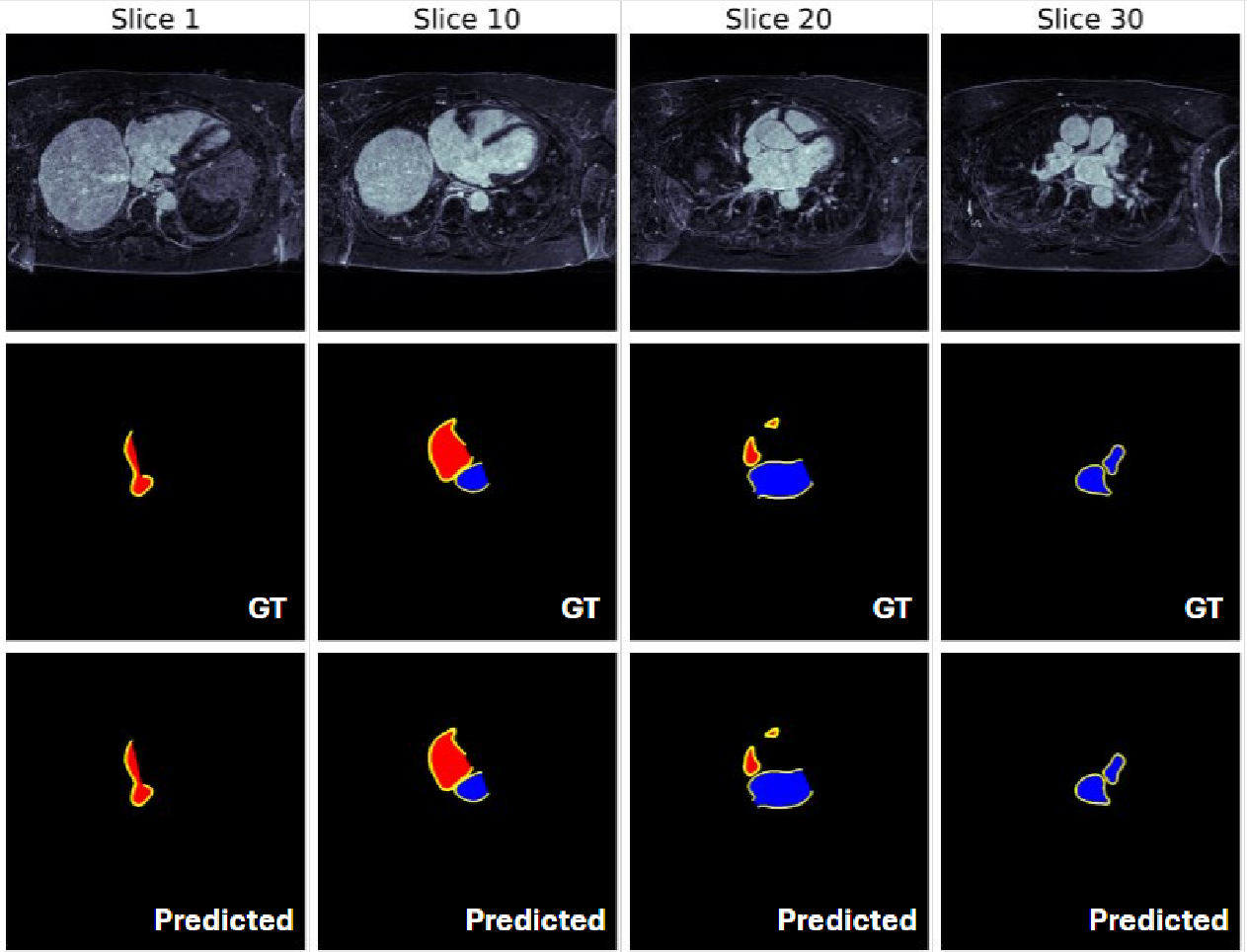}
\caption{A representative example case comparing the ground truth (GT) and predicted masks at every 10 slices. \textbf{(Top)} LGE image (from the training set); \textbf{(Middle)} GT segmentation masks; \textbf{(Bottom)} final predicted masks after post-processing. Key: left atrium (\textcolor{blue}{blue}), right atrium (\textcolor{red}{red}), and myocardial wall (\textcolor{yellow}{yellow}).
}\label{masks_example}
\end{figure}

Our proposed two-stage nnU-Net-based pipeline demonstrates high accuracy in segmenting the LA and RA. The 5-fold cross validation results on the 70-subject training set are: a DSC of $0.92 \pm 0.03$, $0.93 \pm 0.03$ and $0.71 \pm 0.05$, and HD95 of \SI{3.89 \pm 6.67}{\mm}, \SI{4.42 \pm 1.66}{\mm} and \SI{3.94 \pm 1.83}{\mm} for the LA, RA and myocardial wall, respectively. Overall, we see accurate segmentation of the atria, including their appendages. A representative case can be viewed in Fig.~\ref{masks_example}.

\begin{figure}[htb!]
\centering
\includegraphics[width=1.0\textwidth]{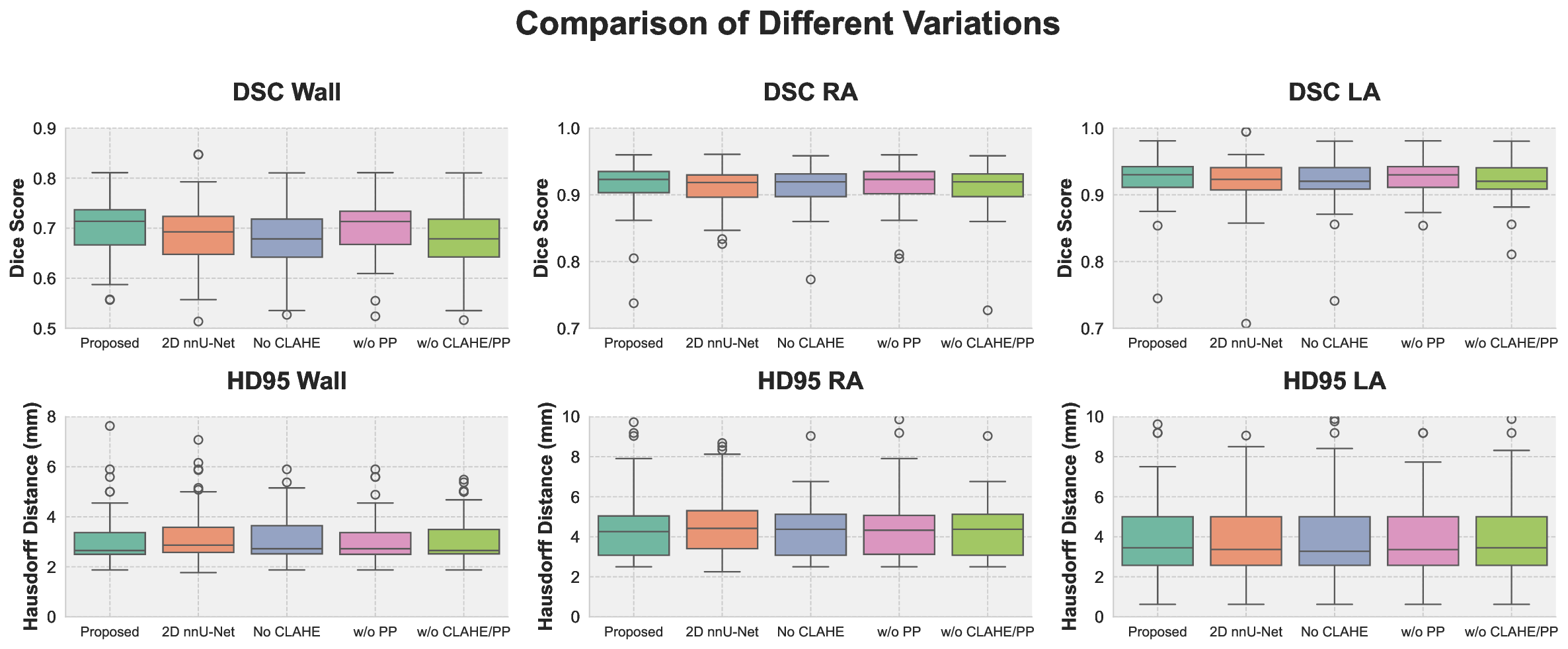}
\caption{The box plots show the Dice similarity coefficient (DSC) \textbf{(Top)} and 95\% Hausdorff distance (HD95) \textbf{(Bottom)} for four other variations of our proposed two-stage model. In each plot, from left to right we show: our proposed method, 2D nnU-Net for segmentation, without CLAHE, without post-processing (PP), and without both CLAHE and PP.}
\label{results}
\end{figure}

Furthermore, we compare our pipeline with four other variants: 1) a 2D nnU-Net for segmentation, 2) without CLAHE, 3) without post-processing, and 4) without both CLAHE and post-processing. As summarised in Fig.~\ref{results}, our proposed pipeline overall outperforms all other variants, with the exception of a drop in performance on the wall's HD95. Specifically, the use of both pre-processing and post-processing yields DSC improvements of 0.8\%, 0.58\% and 4.3\%, and HD95 changes of -32.71\%, -5.22\% and 16.74\% for the LA, RA and wall, respectively. While the 2D nnU-Net variant is advantageous in terms of the execution time, it exhibits a large performance drop compared to its 3D counterpart, especially for the wall, due to limited spatial information.


\begin{figure}[htb!]
\centering
\includegraphics[width=1.0\textwidth]{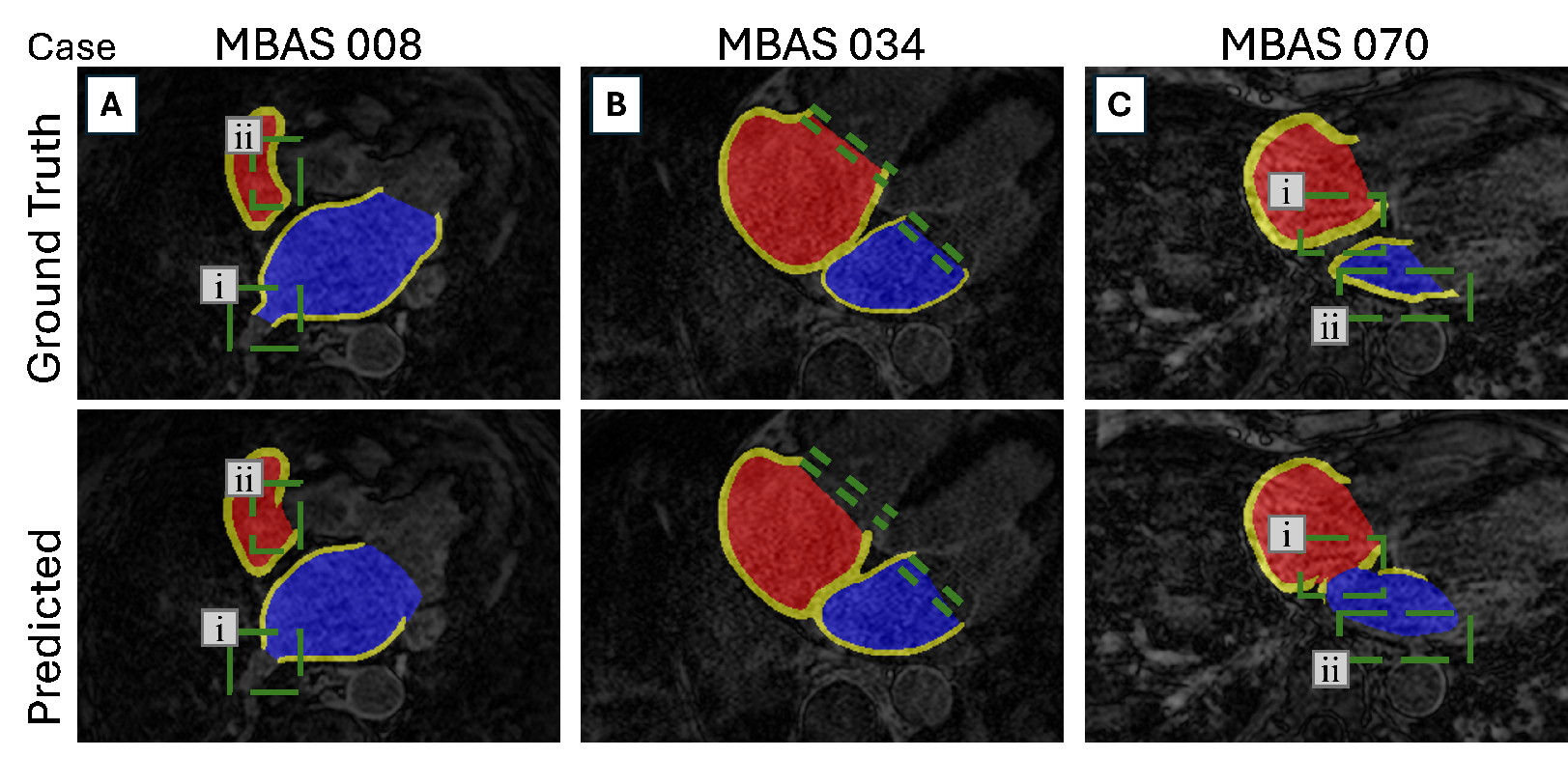}
\caption{Examples of prediction errors from three different cases, with each column representing a separate case. \textbf{Top}: ground truth segmentations; \textbf{Bottom}: predicted segmentations. Marked regions in \textcolor{green}{green} highlight prediction errors.}
\label{errors}
\end{figure}




\section{Discussion}
The proposed pipeline accurately delineates the LA, RA, and myocardial wall from the LGE-MRIs. Our results compare well with reported literature metrics (LA DSC: $0.93\pm0.02$~\cite{Xiong2021AImaging}; RA DSC: $0.90\pm n/a$~\cite{zhu2024ras}) for neural network-based LGE-MRI segmentations. This performance is notable given the computational time and hardware restrictions imposed by the competition: inference times no longer than 10 minutes for 100 test cases on a single Nvidia Tesla V100 GPU and exclusion of additional training data or pre-trained models. We pragmatically focus on the pre-processing and post-processing steps to enhance the segmentation performance with our small training dataset of 70 cases.

In our two-stage approach, cropping plays a crucial role in improving subsequent segmentation inference time by reducing the volume size, while helping to standardise image dimensions. Using the nnU-Net for segmentation is an ideal first choice due to its proven performance and extensive training pipeline~\cite{Isensee2021NnU-Net:Segmentation}. We improve the performance by applying CLAHE before the segmentation, which enhances boundary contrast (Fig.~\ref{results}). In addition, the post-processing steps further refine the segmentation, especially in slices where the masks appear to be more truncated or with blurred boundaries with the adjacent masks.

However, common errors persist, including difficulties in accurately delineating the boundaries between the tricuspid valve and the RA, as well as the mitral valve and the LA, as shown in Fig.~\ref{errors}B. Additionally, structures like the pulmonary veins, superior vena cava, and inferior vena cava are occasionally trimmed (see pulmonary vein example in Fig.~\ref{errors}Ai) or overextended in comparison to the GT. These issues arise due to the lack of distinct features that the model can rely on for precise segmentation. Additionally, these errors are further exacerbated by unavoidable inconsistencies in manual annotation.

As shown in Fig.~\ref{masks_example}, the myocardial wall is accurately delineated around both the RA and LA, with uniform thickness. The RA wall ($\sim$ \SI[separate-uncertainty-units = repeat]{3.54 \pm 0.19}{\mm}) is thicker than the LA wall ($\sim$ \SI[separate-uncertainty-units = repeat]{2.31 \pm 0.23}{\mm}), consistent with the GT. However, small segments of the wall are occasionally missed, either due to low contrast neighbouring tissue being misinterpreted as atrium openings (Fig.~\ref{errors}Aii), or due to high levels of noise and blurring (Fig.~\ref{errors}Cii). Additionally, the model may struggle to accurately identify the thin septal wall separating the LA and RA blood pools, due to its inherently low contrast (Fig.~\ref{errors}Ci). This issue is worsened in images with high noise or ringing artefacts.

Omitting CLAHE results in the largest decline in the wall's DSC to $0.64 \pm 0.56$ and an increase in HD95 to \SI{5.04 \pm 5.15}{\mm}. This highlights CLAHE’s ability to enhance boundary contrast at important regions of the wall, with only negligible computational cost of 0.5~s per cropped case. On the other hand, omitting post-processing has a smaller impact on the overall performance, as expected. Nevertheless, it plays a key role in addressing more challenging cases where noise and artefacts are prominent, preventing large errors in boundary identification and refining segmentations.

The combined results indicate that while CLAHE is essential for improving wall segmentation accuracy, post-processing serves as an important corrective step, especially for outlier cases with high noise or imaging artefacts. However, we note that each of these steps results in small changes, either positive or negative, for the LA and RA segmentations. This might suggest additional benefits in improving the neural network architecture. 


We aim to improve our two-stage nnU-Net model through a multi-task approach \cite{zhao2023multi}, as prior research has shown it can improve robustness and generalisability \cite{zhang2021survey}. Our current model performs poorly on low-quality images, so we plan to develop a joint nnU-Net that integrates segmentation, de-noising, and in-painting to improve both image quality and segmentation accuracy. In addition, we also plan to further improve post-processing steps by using contour correction and wall dilation. Unfortunately, these steps would unavoidably increase inference time, making them unsuitable under the challenge's constraints.

Atrial segmentation is a crucial step that enables quantitative analysis of atrial structure and function. This leads to important applications in prognosis, diagnosis, and treatment. However, the LA and RA are naturally challenging to segment, especially when compared with the left ventricle. This is primarily due to several factors: 1) the complex geometries of the LA and RA, 2) variability in atrial size and shape among individuals, and 3) thin and inconsistent wall thickness~\cite{Barbero2017AnatomyAtria,Ho2009TheFibers,Ho2012LeftRevisited,Varela2015,Varela2017}. Therefore, continuous advancements of automated segmentation methods are needed to further improve the LA and RA analysis. 

\section{Conclusions}
For the MBAS challenge, we propose an automatic pipeline to accurately segment the LA, RA, and myocardial wall from LGE-MRIs. Our segmentation is based on the 3D convolutional nnU-Net and incorporates important pre-processing steps: 2D nnU-Net-based image cropping, CLAHE-based image equalisation, and morphological post-processing steps.

Despite the challenges posed by image quality variations and the complex anatomical geometries of the atria, our approach achieves a competitive DSC and HD95 with a very quick inference time (average 19.5~s). Our pipeline has the potential to support clinical decision-making and improve the planning of treatments like catheter ablation in AF patients.

\textbf{Acknowledegments.}
This work was supported by: St George’s Hospital Charity, the NIHR Imperial Biomedical Research Centre (BRC) and the British Heart Foundation Centre of Research Excellence at Imperial College London (RE/18/4/34215). We acknowledge the computational support provided by the Imperial College Research Computing Service (DOI: 10.14469/hpc/2232).

%
%
%
\bibliographystyle{unsrt}
\bibliographystyle{splncs04}

\bibliography{references_2} 

\end{document}